\documentclass[pre,twocolumn,superscriptaddress,showpacs]{revtex4}

\usepackage{graphicx}
\usepackage{amsmath}
\usepackage{amssymb}

\begin{document}

\title{Towards deterministic equations for L{\'e}vy walks: the fractional
material derivative}
\author{Igor M. Sokolov}
\email{igor.sokolov@physik.hu-berlin.de}
\affiliation{Institut f\"ur Physik, Humboldt-Universit\"at zu Berlin,
Invalidenstra{\ss}e 110, 10115 Berlin, Germany}
\author{Ralf Metzler}
\email{metz@nordita.dk}
\affiliation{NORDITA, Blegdamsvej 17, DK-2100 Copenhagen {\O}, Denmark}

\date{\today}

\begin{abstract}
L{\'e}vy walks are random processes with an underlying spatiotemporal
coupling. This coupling penalizes long jumps, and therefore L{\'e}vy
walks give a proper stochastic description for a particle's motion
with broad jump length distribution. We derive a generalized dynamical 
formulation for L{\'e}vy walks in which the {\em fractional\/} equivalent of the
{\em material derivative\/} occurs. Our approach will be useful
for the dynamical formulation of L{\'e}vy walks in an external force
field or in phase space for which the description in terms of the continuous
time random walk or its corresponding generalized master equation are
less well suited.
\end{abstract}


\maketitle

Anomalous diffusion processes are characterized by deviations from the
traditional linear time dependence $\langle x^{2}(t)\rangle =2Kt$ of the
mean squared displacement in the force-free limit. In particular, one
distinguishes subdiffusion ($0<\alpha <1$) and superdiffusion ($\alpha >1$%
) for the wide class of systems displaying a power-law anomaly $\langle
x^{2}(t)\rangle =2K_{\alpha }t^{\alpha }/\Gamma (1+\alpha )$; here, $%
K_{\alpha }$ is a generalized diffusion constant \cite{bouchaud,hughes}. A
versatile framework for the description of anomalous diffusion are
continuous time random walks (CTRWs), which define a random walk that is
governed by two probability density functions (pdfs), the jump length and
waiting time distributions $\lambda (x)$ and $\psi (t)$ from which the jump
length $x$ and the waiting time $t$ of each jump are drawn \cite{klablushle}%
. Although the stochastic formulation of the CTRW fully defines the random
process and leads to the closed integral equation for the pdf of the
particle's position $P(x,t)$ in terms of $\lambda (x)$ and $\psi (t)$, its
mathematical handling gets awkward as soon as non-natural boundary conditions,
the presence of external force fields, or the description in phase space are
considered. The same complication holds true for the formulation in terms of
generalized master equations, which are equivalent to CTRWs with uncorrelated
$\lambda (x)$ and $\psi (t)$ \cite{gme}. In such cases, the corresponding
deterministic equations of the generalized Fokker-Planck type, in which the
drift terms occur explicitly and which can be attacked with the
standard mathematical tools, render a much more amenable description. To find
such equations for anomalous transport statistics has been a focal point in
stochastic systems studies \cite{report}.

For subdiffusion processes, a complete framework of transport equations has
been established, namely the fractional Fokker-Planck and Klein-Kramers
equations \cite{report,fpe,gcke}. These are natural generalizations of their
Brownian counterparts, and their solution exists whenever the solution of the
corresponding regular Fokker-Planck equation exists, as they correspond to a
\emph{subordination\/} of the analogous normal stochastic process
\cite{report,gcke,igor,eli}.

The description of superdiffusive processes within the same framework is
still far from being completed. Whereas L{\'e}vy flights in the absence of
an external force display a L{\'{e}}vy stable pdf and therefore a diverging
mean squared displacement (and thus could apply only to rather exotic
physical processes) \cite{bouchaud,hughes,klablushle,report,rem}, {\em
L{\'e}vy walks (LWs)\/} give a proper dynamical description in the
superdiffusive domain. The temporal and spatial variables of LWs are strongly
correlated, their steps being governed by a joint distribution $\psi(x,t)$,
in which waiting time and step length pdfs, $\psi(t)$ and $\lambda(x)$ are no
longer independent. In LWs, the occasional long
jumps which are typical for L{\'{e}}vy flights are penalized through the
introduction of a time cost. This spatiotemporal coupling can be achieved,
in the simplest case, by the choice $\psi(x,t)=\frac{1}{2}\psi(t)\delta
(|x|-vt)$, i.e., by a constant velocity \cite{klablushle}. Such a model arises
naturally when describing some limiting cases of molecular collisions
\cite{colli}; its close relatives are reasonable candidates for describing
turbulent dispersion \cite{turbo}.
A question of fundamental interest is therefore the formulation of LWs in
terms of deterministic equations. Whereas previous approaches \cite{basil,msm}
in terms of fractional Klein-Kramers equations could reproduce lower order
moments of an LW, they were hampered by the fact that they could not
describe the full pdf. The main complication on this way is the fact, that
the overall LW process cannot be immediately considered as subordinated
to a Wiener one (or to a simple random walk); however, as we proceed to show,
it is exactly the strong correlation of the temporal and the spatial aspects
of LWs which makes it possible to provide a description based on a process
subordinated to a simple two-state Markovian process. In the present work, we
derive the \emph{exact\/} deterministic evolution equation for LWs which
holds both for the free motion and in a constant force field. In the
following, we use rescaled quantities and concentrate on the one dimensional
case.

We first define a two-state Markovian random process describing the velocity
switching and then proceed to generalize it to two different domains of LWs.
Thus, let us denote by $P_{+}$ and $P_{-}$ the probabilities to move to the
right or to the left, respectively. Probability conservation demands that $%
P_{+}+P_{-}=1$. Moreover, for simplicity we assume that the absolute value
of the velocity of motion to the right and to the left is 1. Within the rate
description, for a symmetric case, the probabilities $P_{\pm }$ satisfy the
differential equation 
\begin{equation}
\frac{d}{dt}P_{\mp }=P_{\pm }-P_{\mp }  \label{Regular}
\end{equation}
Eq.~(\ref{Regular}) can be readily solved: Taking $P_{-}=1-P_{+}$ we get: $%
\frac{d}{dt}P_{+}=1-2P_{+}$. The equilibrium situation corresponds to $%
P_{+}=P_{-}=1/2$, and the relaxation to this equilibrium from the initial
condition $P_{+}=1$ is exponential, 
\begin{equation}
P_{\pm }={\textstyle \frac{1}{2}\pm \frac{1}{2}}\exp (-2t).  \label{Solution}
\end{equation}

Let us concentrate first on the switching process described by this
equation: It is an alternating random process with the waiting time pdf $\psi
(t)=\exp (-t).$ At each ''tick'' the state is changed from $+1$ to $-1$ and
back. $P_{+}(t)$ is then the probability that at time $t$, the state of the
system is ''$+1$'', i.e. that the overall number of full steps (changes of
sign) was even. The Laplace transform of this probability is 
\begin{eqnarray}
P_{+}(u) &=&\sum_{n=0}^{\infty }\chi _{2n}(u)=\frac{1-\psi (u)}{u}%
\sum_{n=0}^{\infty }\psi ^{2n}(u)  \nonumber \\
&=&\frac{1}{u\left[ 1+\psi (u)\right] }.  \label{Peplus}
\end{eqnarray}
A similar expression for $P_{-}$ involves the summation over the odd
numbers of steps. In Eq.~(\ref{Peplus}), the $\chi
_{2n}(t)$ denote that the walker performs $2n$ direction changes within an
overall waiting time $t$. The latter is denoted by $\Psi
(t)=1-\int_{0}^{t}\psi (t)dt$ with Laplace transform $\psi (u)=u^{-1}-\psi
(u)/u$ \cite{klablushle}. For our exponential function ($\psi (u)=1/(1+u)$),
the result becomes $P_{+}(u)=\frac{1}{2u}+\frac{1}{2(2+u)}$, which is
exactly the Laplace transform of Eq.~(\ref{Solution}).

Consider now a a long-tailed waiting time pdf $\psi (t)\sim t^{-1-\alpha }$
of the explicit form \cite{rem1} 
\begin{equation}
\psi (u)=\frac{1}{1+u^{\alpha }},\quad 0<\alpha <1  \label{7}
\end{equation}
in Laplace space. This specific form has the following origin due to
subordination \cite{SokSub}. In a system
whose relaxation in its operational time is given by an exponential 
$\phi(\tau )=\exp (-\tau)$, we can introduce a coarse-graining in
which the operational time is divided into intervals $\Delta \tau $ (and $%
\Delta \tau $ taken as a new time unit). Assume that the duration of the
physical time interval corresponding to $\Delta \tau $ is given by a
one-sided L\'{e}vy distribution. The duration of the physical time
corresponding to the interval $\tau $ is then a convolution of $n=\tau
/\Delta \tau $ such distributions, and its Laplace transform is $\exp
(-nu^{\alpha })$. Averaging over $n$ we get $\psi (u)\sim \int \exp
(-nu^{\alpha })\exp (-n)dn$, exactly reproducing Eq.~(\ref{7}). As an
example we can explicitly determine $\psi (t)$ for $\alpha=\frac{1}{2}$:
$\psi_{1/2}(t)=(\pi t)^{-1/2}-e^{t}\mbox{erfc}\sqrt{t}$, with the asymptotic
behaviors $t^{-1/2}$ ($t\ll 1$) and $t^{-3/2}$ ($t\gg 1$).

With the waiting time pdf (\ref{7}), Eq.~(\ref{Regular}) generalizes to the
fractional form 
\begin{equation}
\frac{d}{dt}P_{\mp}=\,_{0}D_{t}^{1-\alpha }\Big(P_{\pm}-P_{\mp}\Big)
\label{3}
\end{equation}
where $_0D_t^{1-\alpha}=\frac{d}{dt}\,_{0}D_{t}^{1-\alpha}$, and $_0D_t^{
-\alpha}$ is the fractional Riemann-Liouville integral operator defined in
terms of 
\begin{equation}  \label{rl}
_0D_t^{-\alpha}f(t)\equiv\frac{1}{\Gamma(\alpha)} \int_0^t
dt^{\prime}f(t^{\prime})(t-t^{\prime})^{\alpha-1}
\end{equation}
with the convenient property $\int_0 ^{\infty}e^{-ut}\,_0D_t^{-\alpha}f(t)=
u^{-\alpha}f(u)$ \cite{report,oldham}. In Eq.~(\ref{3}), the fractional
derivative on the rhs describes a process which is subordinated to the
simple exponential switching under the operational time given by the L{\'e}%
vy stable waiting time pdf. To see this let us compare two solutions: one
using the ''CTRW''-time and another solving Eq.~(\ref{3}) directly:

From Eq.~(\ref{3}), with the initial conditions $P_{+}(0)=1$ and $P_{-}(0)=0$,
we recover upon Laplace transformation 
\begin{eqnarray*}
uP_{+}-1 &=&-u^{1-\alpha }P_{+}+u^{1-\alpha }P_{-} \\
uP_{-} &=&u^{1-\alpha }P_{+}-u^{1-\alpha }P_{-};
\end{eqnarray*}
from the second equation, $P_{-}=\frac{1}{u^{\alpha }+1}P_{+}$, and
therefore we find by insertion into the first: 
\begin{equation}
P_{+}=\frac{1+u^{\alpha }}{2u+u^{\alpha +1}},\mbox{ and }P_{-}=\frac{1}{%
2u+u^{\alpha +1}}.  \label{lapform}
\end{equation}
It is easy to verify that the same result is obtained by combining Eq.~(\ref
{7}) with Eq.~(\ref{Peplus}). Eq.~(\ref{lapform}) describes
the kinetics of moving to the left and to the right. 

We now combine the
purely temporal results for $P_{\pm }$ with the drift invoked by a constant
velocity, distinguishing between two different cases. The ensuing propagator
$P$ of the associated symmetric random walk is combined from a
superposition of two realizations of the switching process, 
taking place with the rate of 1/2 each, one in which the first step 
goes to the right, and one in which it goes left. 
This introduces an additional factor 1/2 in all the following equations.  

\emph{(i) Ballistic regime.} In the Markovian case, the combination of the
process (\ref{Solution}) with a velocity of magnitude 1 introduces the
material derivatives $d_{\pm }\equiv \frac{\partial }{\partial t}\pm \frac{%
\partial }{\partial x}$. Viewing now $P_{\pm }$ as functions of $x$ and $t$,
the evolution equation for $P_{\pm }(x,t)$ result: 
\begin{equation}
d_{\pm }P_{\pm }={\textstyle\frac{1}{2}}(P_{\mp }-P_{\pm }).
\label{Material}
\end{equation}
Both together produce the telegrapher's equation 
\begin{equation}
\frac{\partial }{\partial t}P+\frac{\partial ^{2}}{\partial t^{2}}P=\frac{%
\partial ^{2}}{\partial x^{2}}P  \label{telegraph},
\end{equation}
also known as Cattaneo equation \cite{cattaneo}, for the quantity of interest,
the propagator $P=P_{+}+P_{-}$. The Cattaneo equation describes a process
which at short times behaves ballistically, $\langle x^{2}(t)\rangle\sim t^{2}$,
and at long times exhibits normal diffusion, $\langle x^{2}(t)\rangle\sim t$.
This derivation was based on the material derivatives $d_{\pm }$, whose
Fourier-Laplace transform is $u\pm ik$. We now demonstrate that we reproduce
exactly the propagator of an LW if we assume ({\em ad hoc\/}) that the
corresponding fractional material derivative is defined through 
\begin{equation}
\mathcal{F}_{x}\mathcal{L}_{t}\left\{ d_{\pm }^{1-\alpha }f(x,t)\right\}
=(u\pm ik)^{1-\alpha }f(k,u).
\label{FuLa}
\end{equation}
This choice is motivated by the fact that a waiting time is still coupled to
a walk of length $x=t$, and that for anomalous transport processes the
Fourier-Laplace space is the natural basis to introduce generalizations.
Thus, we obtain 
\begin{equation}
d_{\pm }P_{\pm }=\frac{1}{2}d_{\pm }^{1-\alpha }\Big(P_{\mp }-P_{\pm }\Big),
\label{Frak}
\end{equation}
where the fractional material derivatives are to be interpreted in terms of (%
\ref{FuLa}). We solve Eq.~(\ref{Frak}) under the initial condition $%
P_{+}(x,0)=P_{-}(x,0)=\delta (x)/2$ so that, when introducing the propagator 
$P=P_{+}+P_{-}$ and its counterpart $Q=P_{+}-P_{-}$, we have $P(x,0)=\delta
(x)$ and $Q(x,0)=0$. With the abbreviations $\lambda _{+}=u+ik$ and $\lambda
_{-}=\lambda _{+}^{*}=u-ik$, Eqs.~(\ref{Frak}) can be rewritten in terms of
Fourier-Laplace transformed $P$ and $Q$, as 
\begin{equation}
\frac{1}{2}\lambda _{\pm }(P\pm Q)-\frac{1}{2}=\mp \frac{1}{2}\lambda
_{\pm}^{1-\alpha }Q  \label{frama}
\end{equation}
The solution for $P$ reads: 
\begin{equation}
P=\frac{\lambda _{+}^{\alpha }\lambda _{-}^{\alpha -1}+\lambda _{-}^{\alpha
}\lambda _{+}^{\alpha -1}+\lambda _{+}^{\alpha -1}+\lambda _{-}^{\alpha -1}}{%
\lambda _{+}^{\alpha }+\lambda _{-}^{\alpha }+2\lambda _{+}^{\alpha }\lambda
_{-}^{\alpha }}.  \label{Result}
\end{equation}
Let us show that this is an exact expression for the LW with the waiting
time pdf (\ref{7}). To this end, note that within the CTRW the propagator is
obtained as $P(k,u)=\Psi (k,u)/[1-\psi (k,u)]$ in the case of spatiotemporal
coupling with $\psi (x,t)=\frac{1}{2}\left[ \delta (x-t)+\delta (x+t)\right]
\psi (t)$ and $\Psi (x,t)=\frac{1}{2}\left[ \delta (x-t)+\delta (x+t)\right]
\Psi (t)$ \cite{zukla}. Consequently, we find 
\[
\psi (u,k)=\frac{1}{2}\left[ \psi (u+ik)+\psi (u-ik)\right] 
\]
and an analogous expression for $\Psi (u,k)$, such that we arrive at the
Fourier-Laplace form of $P$: 
\begin{equation}
P=\frac{\left[ 1-\psi (\lambda _{+})\right] /\lambda _{+}+\left[ 1-\psi
(\lambda _{-})\right] /\lambda _{-}}{2-\psi (\lambda _{+})-\psi (\lambda
_{-})}.  \label{CTRW}
\end{equation}
With $\psi (u)$ given by Eq.~(\ref{7}), Eq.~(\ref{Result}) is reproduced and
we have shown that Eq.~(\ref{frama}) with the definition (\ref{FuLa})
describes an LW. From the representation (\ref{CTRW}), we find the Laplace
space form $\langle x^{2}(u)\rangle =2\left( u^{\alpha }+1-\alpha \right)
/\left( u^{3}+u^{3+\alpha }\right) $ of the second moment, from which we
obtain the limiting behaviors $\langle x^{2}(t)\rangle \sim t^{2}$ for $t\ll 1
$ and $\langle x^{2}(t)\rangle \sim (1-\alpha )t^{2}$ for $t\gg 1$, i.e., a
mere decrease in the amplitude of an \emph{overall ballistic\/} process: the
memory which we introduced by the long-tailed form of $\psi $ leads to an
extreme persistence in a given direction on all time scales. There is no
turnover to a process with a smaller exponent of $t$ as in the Cattaneo case.

\emph{(ii) Subballistic regime.} Let us now compare this with the better known
case of the subballistic domain. We again follow our above obtained recipe
of formulating two equations for the direction-switching process with the
waiting-time distribution of interest, and then change the time derivatives
for the material ones. There is an heuristic way immediately leading to the
equations: in the Laplace representation, the system of equations 
\begin{eqnarray}
uP_{+}-1 &=&{\textstyle\frac{1}{2}}f(u)(-P_{+}+P_{-})  \nonumber \\
uP_{-} &=&{\textstyle\frac{1}{2}}f(u)(P_{+}-P_{-})
\end{eqnarray}
leads to the solution $P_{+}=\frac{2(u+f(u))}{u(u+2f(u))}$. Noting that
according to Eq.~(\ref{Peplus}), this is $\left\{ u\left[ 1+\psi (u)\right]
\right\} ^{-1}$, we find the relation $\psi (u)=\frac{1}{1+u/f(u)}$. If we
want a function behaving for small $u$ as $\psi(u)\sim 1-u-u^{1+\beta}$ (i.e.,
one with $\alpha =1+\beta $ being in the interval between 1 and 2) we have
to choose $f(u)=1+u^{\alpha }$, which produces $\psi (u)=\frac{1+u^{\beta }}{
1+u^{\beta }+u}$. This corresponds to the equations 
\begin{equation}
\frac{d}{dt}P_{\pm }=\frac{1}{2}\left( 1+\,_{0}D_{t}^{\beta }\right) \Big(%
P_{\mp }-P_{\pm }\Big)
\end{equation}
for the alternating process, and to the equations with material derivatives 
\begin{equation}
d_{\pm }P_{\pm }=\frac{1}{2}\Big(1+d_{\pm }^{\beta }\Big)\Big(P_{\mp
}-P_{\pm }\Big)
\label{FuLa1}
\end{equation}
for the L\'{e}vy walk. Using our formal rules, we get in the Fourier-Laplace
representation: 
\[
\frac{1}{2}\lambda _{\pm }(P\pm Q)-\frac{1}{2}=\mp \frac{1}{2}(1+\lambda
_{+}^{\beta })Q,
\]
from which we find the result: 
\begin{equation}
P=\frac{2+\lambda _{+}^{\beta }+\lambda _{-}^{\beta }+\lambda _{+}+\lambda
_{-}}{2\lambda _{+}\lambda _{-}+\lambda _{+}(1+\lambda _{-}^{\beta
})+\lambda _{-}(1+\lambda _{+}^{\beta })}.
\end{equation}
This corresponds exactly to Eq.~(\ref{CTRW}) for the new $\psi $ and again
corroborates the recipe to generalize $d_{\pm }$ to the fractional material
derivatives $d_{\pm }^{\beta }$ in the Fourier-Laplace domain. The second
moment of this process is obtained as $\langle x^{2}(u)\rangle =2\left(
u+\beta u^{\beta }\right) /\left( u^{4}+u^{3+\beta }+u^{3}\right) $,
producing the limiting behaviors $\langle x^{2}(t)\rangle \sim t^{2}$ for $%
t\ll 1$ and $\langle x^{2}(t)\rangle \sim 2\beta t^{2-\beta }/\Gamma
(3-\beta )$ for $t\gg 1$, i.e., a transition from initial ballistic to
terminal subballistic superdiffusive behavior, in analogy to the CTRW
result \cite{klablushle,zukla}.

Let us now discuss the coordinate-time representation of the fractional
material derivative. Using the standard relation $F(u+b)\fallingdotseq
e^{-bt}f(t)$ of the Laplace transformation we obtain after some steps 
\begin{equation}
d_{\pm }^{\alpha }P(x,t)=\,_{0}D_{t}^{\alpha }P(x\pm t,t).
\end{equation}
The fractional material derivative, this is, generalizes the standard
material derivative, $d_{\pm}P(x,t)=\frac{d}{dt}P(x\pm t,t)\equiv\left( 
\frac{\partial}{\partial t}\pm\frac{\partial}{\partial x}\right)P$ for $%
\alpha=1$, through the introduction of the standard Riemann-Liouville
operator acting on the entire right hand side.

For subdiffusion, the major advantage of the fractional dynamical equation
formulation is in the possibility to easily generalize to situations with an
external force field, which led to the fractional Fokker-Planck equation 
\cite{report,fpe}. Here, we start with incorporating a constant external
force. To this end, let us consider the physical realization of the walk in a
splitting flow: in the upper half-plane, the particle moves to the right; $%
v_{x}^{+}=v_{0}$ for $y>0$, in the lower half-plain it moves to the left: $%
v_{x}^{-}=-v_{0}$ for $y<0$. The motion in the $y$-direction dictates the
waiting-time distribution. If it is simple diffusion, the overall process is
a L\'{e}vy-walk with $\alpha =1/2$ \cite{bouchaud}. Imagine now, we have a
force acting in
the $x$-direction. The force causes sliding of the particle with respect to
the flow, so that now $v_{x}^{+}=v_{0}+\mu f$ for $y>0$ and $%
v_{x}^{-}=-v_{0}+\mu f$ for $y<0$. This corresponds to changing $d_{\pm
}^{\alpha }$ to the constructs corresponding to 
\begin{equation}
d_{f,\pm }^{\alpha }P(x,t)=\,_{0}D_{t}^{\alpha }P(x+[\mu f\pm v]t,t),
\end{equation}
whose Fourier-Laplace transform produces 
\begin{equation}
d_{f,\pm }^{\alpha }\rightarrow \left[ u+i(\mu f\pm v)k\right] ^{\alpha }.
\end{equation}
That is, in both the force-free and the constant force cases, we observe
some type of generalized d'Alembert principle reflecting the $\delta$-coupling
of $x$ and $t$ \cite{rem2}.

Translating the dynamic equations (\ref{FuLa}) and (\ref{FuLa1}) with the
fractional material derivatives into an equation for the propagator
$P$ produces a rather complicated expression. This can in fact be circumvented
by a somewhat different definition of the fractional operators, as shown in
Ref.~\cite{meerschaert}. However, the latter does not allow for the
incorporation of a bias and is therefore not suited for our purpose.

In our approach we were guided by the equivalence between position $x$ and
time $t$ in the LW framework, enforced by the $\delta $-coupling $\psi (x,t)=%
\frac{1}{2}\delta (|x|-t)\psi (t)$, which could in fact be rewritten in
terms of the jump length distribution as $\frac{1}{2}\delta (|x|-t)\lambda
(x)$, with the appropriate long-tailed form for $\lambda $. This equivalence
gives rise to the occurrence of the material derivative, in complete analogy
to the Brownian Cattaneo case. However, in the presence of long-tailed
temporal correlations of the kind $\psi (t)\sim t^{-1-\alpha }$, the
fractional variant of the material derivative emerges, with its simple
representations in both Fourier-Laplace and $(x,t)$ domains. This treatment
is amenable for the case of constant external force. Whether there is a
similar treatment for arbitrary force $f(x)$ is not clear at present.


\begin{thebibliography}{99}
\bibitem{rem1} This is the Laplace transform of a pdf, since $\psi(0)=1$
and $\psi(u)$ is a completely monotone function: it has a form $g(f(u))$,
where $g(s)=1/s$ is a completely monotone function (it is positive, and
the signs of its derivatives alternate), and $f(u)=1+u^{\alpha}$ is a
function with completely monotone derivative $f^{\prime }(u)=\alpha
u^{\alpha -1}$.
\bibitem{rem2} For a general external force field $f(x)$, the expected form
of the fractional material derivative is expected to read
\begin{eqnarray*}
d_{f(x),\pm }^{\alpha }P(x,t)&=&\,_{0}D_{t}^{\alpha }P\left( x(t),t\right)\\
&=&\,_{0}D_{t}^{\alpha }P\left( x\pm vt+\int_{0}^{t}\mu f(x(t))dt,t\right);
\end{eqnarray*}
however, there is no direct Fourier-Laplace transform of this operator.
\end{thebibliography}
\end{document}